\documentclass[amsmath,reprint,amssymb,aps,prd]{revtex4-1}

\usepackage{graphicx}
\usepackage{dcolumn}
\usepackage{bm}
\usepackage{hyperref} 
\hypersetup{linkcolor=red,citecolor=blue,filecolor=dullmagenta,urlcolor=darkblue}

\newcommand{\avg}[1]{\langle #1\rangle}

\begin{document}

\title{Cosmological backreaction in the presence of radiation and a cosmological constant}

\author{Viraj A. A. Sanghai}
\email{v.a.a.sanghai@qmul.ac.uk}
\author{Timothy Clifton}
\email{t.clifton@qmul.ac.uk}
\affiliation{School of Physics and Astronomy, Queen Mary University of London, London E1 4NS,
UK.}

\begin{abstract}
We construct high-precision models of the Universe that contain radiation, a cosmological constant, and periodically distributed inhomogeneous matter. The density contrasts in these models are allowed to be highly non-linear, and the cosmological expansion is treated as an emergent phenomenon. This is achieved by employing a generalised version of the post-Newtonian formalism, and by joining together inhomogeneous regions of space-time at reflection symmetric junctions. Using these models, we find general expressions that precisely and unambiguously quantify the effect of small-scale inhomogeneity on the large-scale expansion of space (an effect referred to as ``back-reaction", in the literature). We then proceed to specialize our models to the case where the matter fields are given by a regular array of point-like particles. This allows us to derive extremely simple expressions for the emergent Friedmann-like equations that govern the large-scale expansion of space. It is found that the presence of radiation tends to reduce the magnitude of back-reaction effects, while the existence of a cosmological constant has only a negligible effect.
\end{abstract}


\maketitle

\section{Introduction}

In previous work we developed a new formalism for constructing cosmological models with a periodic lattice structure \cite{vaas, Tim1}. This was done by taking regions of space-time that we described using the post-Newtonian perturbative expansion, and patching them together at reflection symmetric boundaries to form a global solution to Einstein's equations. The advantages of this approach are (i) that it allows extremely large density contrasts to be consistently included in cosmology, at higher orders in perturbation theory, without the imposition of any continuous symmetries (i.e. Killing vectors), and (ii) that it allows the cosmological expansion to be viewed as an emergent phenomenon, resulting from the junction conditions between patches \cite{Is1}, rather than being specified from the outset. 

These two features, taken together, make our lattice models ideally suited for studying the effects that non-linear structure has on the large-scale expansion of space. Such effects, usually referred to as ``back-reaction" in the cosmology literature \cite{br1,br2,br3}, are important to understand if we are to have faith that the homogeneous and isotropic Friedmann-Lema\^{i}tre-Robertson-Walker (FLRW) models are suitable for interpreting observations in a lumpy universe (such as the one within which we live). They may also be important for the much heralded era of ``precision cosmology"  \cite{euclid, ska}, especially if observations become good enough to isolate higher-order relativistic effects.

However, while they may constitute interesting devices for studying back-reaction, and while they can help to illustrate the complementary nature of cosmology and weak-field gravity, the lattice models constructed in \cite{vaas, Tim1} are not fully realistic. One way in which this situation can be improved upon, and on which we focus in this paper, is by adding other types of matter fields, beyond the non-relativistic matter that is usually included in studies of post-Newtonian gravity. In this regard, particular matter fields that are of interest in cosmology are radiation, and the cosmological constant, $\Lambda$. The former of these becomes increasingly important at early times, while the latter (if it is non-zero) comes to dominate the expansion at late times.

In this paper we extend the post-Newtonian formalism by including the contribution of barotropic fluids with non-vanishing pressure, $p=p(\rho)$, to the energy-momentum tensor. Such an approach can be used to include a fluid of radiation, with $p=\frac{1}{3} \rho$, or a cosmological constant, with $p=-\rho$. It could also be used to include a variety of other matter fields that are commonly considered in cosmology. We then use this extended formalism to model the gravitational fields that exist within each of our lattice cells, and proceed to determine (lengthy) general expressions for the effect that such fluids have on the large-scale expansion of space. This is done in full generality, without assuming anything about the distribution of matter within each cell.

\begin{figure*}[t]
\includegraphics[width=0.95 \textwidth]{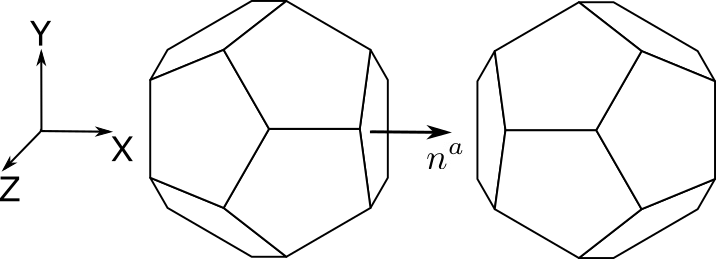}
\caption{\label{fig1} Two neighbouring cells with a dodecahedral shape, that are reflection symmetric around their common cell face, and that form a part of a larger periodic lattice structure. The unit vector, $n^{a}$, is the space-like normal to the boundary of the cell.}
\end{figure*}

In order to develop these ideas further we then specialize the distribution of matter to a particularly simple example: a single point-like mass at the centre of each cell, in the presence of radiation and $\Lambda$. Globally, this corresponds to a regular array of massive particles sitting in a sea of radiation. The result of considering this specific set-up is an expression for cosmological back-reaction that takes an extremely simple form. Its effect on the Friedmann equation is to add an extra term, so that we have
\begin{align}
\left( \frac{\dot{a}}{a} \right)^2 =& \frac{8\pi G}{3}( \rho_{M} + \rho_{r}) -\frac{k}{a^2} +\frac{\Lambda}{3} +\mathcal{B}  \ , \nonumber
\end{align}
where the back-reaction term, $\mathcal{B}$, is given by
\begin{align} 
\mathcal{B} 
&\simeq -\left( 2\pi G L \rho_{M} a \right)^2 \left(1.50 -  0.80 \frac{\Omega_{r}}{\Omega_{M}} + 1.76 \frac{\Omega_{k}}{\Omega_{M}} \right)  \, , \nonumber
\end{align}
where $\Omega_{M}$, $\Omega_{r}$ and $\Omega_{k}$ are the standard cosmological density parameters for matter, radiation and spatial curvature, respectively, and where $L$ is the length of the edge of a cell (see Section \ref{results} for details). 

It can be seen that the discretely distributed matter contributes a term that looks like radiation to the effective Friedmann-like equation that governs the large-scale expansion of space, just as was found in \cite{vaas}. We find that presence of radiation, however, reduces the magnitude of this back-reaction term, while the presence of $\Lambda$ has no noticeable effect on it at all. As shown in \cite{vaas}, a negative value of spatial curvature increases the amplitude of the back-reaction, and a positive value decreases it.

The physical set-up that we consider in the latter parts of this paper, consisting of a universe full of point sources, has received considerable attention over the past few years. This includes studies of the initial data of such models \cite{crt,yoo1,ben1,kor1,kor2}, as well as their evolution \cite{ben2,ben3,yoo2,numlam,cgr,cgr2,hinder}. Studies of back-reaction in the presence of radiation and $\Lambda$ have also been performed using both perturbative methods \cite{lam, radlam1,radlam2}, and by solving the full Einstein equations \cite{numlam, numrad}. Our work is complementary to these previous studies. It builds on them by developing and applying a versatile perturbative framework that incorporates non-linear density contrasts, while avoiding the ambiguities that can arise when averaging in general relativity.

The plan for the rest of this paper is as follows: In section \ref{sec2} we set out the equations that describe the geometry and dynamics of our lattice cells. In section \ref{sec3} we use these equations to determine the cosmological expansion of our lattice, in the presence of an arbitrary barotropic fluid, and for any general distribution of matter. In section \ref{sec4} we then look at the specific case of regularly arranged point masses in cubic cells in the presence of radiation, spatial curvature and a cosmological constant. Throughout the paper we use latin letters ($a$, $b$, $c$, ...) to denote space-time indices, and greek letters ($\mu$, $\nu$, $\rho$, ...) to denote spatial indices. We reserve the first half of the capital latin alphabet ($A$, $B$, $C$, ...) to denote the spatial components of tensors in the boundary of a cell, and the latter half ($I$, $J$, $K$, ...) as labels to denote quantities associated with our various different matter fields.

\section{The geometry of a lattice cell} \label{sec2}

In this section we present the equations that describe the geometry within each of our lattice cells, and the dynamics of their boundaries. We begin by briefly recapping the set-up of our bottom-up approach to cosmology, before moving on to discuss how we extend the post-Newtonian formalism to include a barotropic fluid, as well as non-relativistic matter. After this, we make use of reflection symmetric junction conditions to find the evolution of the boundary of every cell. Altogether, this gives us just enough information to work out the expansion of each of our cells, and hence the lattice as a whole, to the first post-Newtonian level of accuracy.

\subsection{Lattice structure}

We begin by splitting the Universe into a large number of identical cells, in order to construct a periodic lattice structure. When doing this, we allow the cell shapes to be chosen as any regular convex polyhedra that tessellates a three-dimensional space that is either flat or has constant positive or negative spatial curvature. There are six such tessellations for a space of positive curvature, one for a flat space, and four for a space of negative curvature (see Table I of \cite{vaas}, and reference \cite{poly}, for details). The analysis we present in this section is valid for any cell shape, and for any of these possible tessellations.

Let us now consider any one cell, and take, as a first approximation, the space-time within this cell to be close to Minkowski space. We can then choose a Cartesian set of spatial coordinates ($x$, $y$, $z$), and rotate these coordinates until the vector $\partial/ \partial x$ is orthogonal to one of the faces (as all cell faces are identical, it does not matter which one we choose). This situation is illustrated in Fig. \ref{fig1}, for the case of a dodecahedral cell. The position of every point on this cell face is then given by $x = X(t,y,z)$. 

Now, as the evolution of this cell face is a $(2+1)$-dimensional time-like hypersurface, we can define a space-like unit vector, $n^{a}$, as its normal. The leading-order contributions to the covariant components of this vector are given by \cite{vaas}
\begin{equation}
n_a = \left( -X_{,t}, 1, -X_{,y}, -X_{,z} \right) \, ,
\end{equation}
where commas denote partial differentiation. Reflection symmetry implies that junction conditions at the cell face should be invariant under the change $n^a \rightarrow - n^a$, which means that the extrinsic curvature of every $(2+1)$-dimensional cell face must vanish. This requirement provides the information necessary for specifying the boundary conditions for the field equations within each cell, as well as for the motion of the cell face itself.


\subsection{Post-Newtonian expansion}

The matter content and geometry within each of our cells is described using the post-Newtonian perturbative expansion. This formalism, valid in the limit of weak gravitational fields and slow motions, assigns orders of smallness to quantities in the metric and the energy-momentum tensor using the parameter
\begin{equation}  \label{2}
\epsilon \equiv \frac{|\bm{v}|}{c} \ll 1 \, , 
\end{equation} 
where $\bm{v}$ is the  three-velocity associated with the matter fields, and $c$ is the speed of light. The post-Newtonian expansion also requires that time derivatives add an order of smallness, compared to space derivatives, so that $X_{,t} /X_{,y} \sim O(\epsilon)$, {\it etc}.

The explicit expansion of the metric is then given by the following line element:
\begin{align}
ds^2 = \left( -1+ h^{(2)}_{tt} + h^{(4)}_{tt} \right) dt^2  &+ 2 h^{(3)}_{t\mu} dt dx^{\mu} 
\nonumber \\ &\hspace{-10pt} + \left( \delta_{\mu\nu}+ h^{(2)}_{\mu\nu} \right)  dx^{\mu} dx^{\nu} 
\label{metric} \, ,
\end{align}
where $h^{(2)}_{tt}$, $h^{(2)}_{\mu\nu}$, $h^{(3)}_{t\mu}$ and $h^{(4)}_{tt}$ are perturbations to the Minkowski metric, and where superscripts in brackets represent the order of smallness of a quantity. The metric has been expanded to $O(\epsilon^4)$ in the time-time component, to $O(\epsilon^3)$ in the time-space components, and $O(\epsilon^2)$ in the space-space components. These are the orders of accuracy required in order to consistently write the equations of motion for time-like particles to first post-Newtonian accuracy.

We can similarly expand the matter fields in powers of $\epsilon$. To do so, we define the energy density, $\rho$, and isotropic pressure, $p$, as
\begin{align}
\rho =& T_{ab} u^{a} u^{b} \, ,  \\
p =& \frac{1}{3}T_{ab} (g^{ab} + u^{a} u^{b}) \, , 
\end{align}
where $T_{ab}$ is the energy-momentum tensor, $g_{ab}$ is the metric of space-time, and $u^{a}$ is a reference four-velocity that satisfies $u^{a}u_{a} = -1$. We can expand the energy density and pressure as
\begin{align}
\rho =& \rho^{(2)} +\rho^{(4)}+ O(\epsilon^6) \, ,  \\
p =& p^{(2)} + p^{(4)} + O(\epsilon^6) \, , \label{denpress}
\end{align}
and write the expanded four-velocity as
\begin{align} 
u^{a} =&  \bigg(1 +\frac{h^{(2)}_{tt}}{2}+ \frac{v^2}{2}\bigg)(1;v^{\mu}) +  O(\epsilon^4) \, ,
\end{align} 
where $v$ is the three-velocity of the fluid we are considering, and $v^2= v^{\mu}v_{\mu}$.

The reader may note that we have included a contribution to the pressure at $O(\epsilon^2)$, which is usually taken to vanish in post-Newtonian gravity. We have done this in order to include barotropic fluids, which generally have the leading-order contribution to pressure at the same order as energy density. For further details of post-Newtonian expansions, the reader is referred to \cite{will}.
 
\subsection{Matter content} \label{conserve}
 
Let us now consider the matter content of our space-time. We wish to model a universe that contains both non-relativistic matter, with $p^{(2)}=0$, and a barotropic fluid, with equation of state $p=p(\rho)$. For simplicity, and as a first approximation, we will take the latter of these to be a perfect fluid that does not strongly interact with the non-relativistic matter. Such a fluid could be used to model radiation ($p=\frac{1}{3} \rho$), vacuum energy ($p=-\rho$), or a massless scalar field ($p=\rho$). The non-relativistic matter is intended to represent both baryonic matter and cold dark matter.
 
We therefore write the total energy-momentum tensor for these two fluids as
\begin{align}
T^{ab} = T^{ab}_M + T^{ab}_I\, ,
\end{align}
where subscripts $M$ and $I$ refer to quantities associated with the non-relativistic matter fields and the barotropic fluid, respectively. In what follows, the cosmological constant, $\Lambda$, is included directly in the field equations. If we now take the reference four-vector for each of the fluids to be given by
\begin{align} 
u_{M}^{a} =&  \bigg(1 +\frac{h^{(2)}_{tt}}{2}+ \frac{v_{M}^2}{2}\bigg)(1;v_{M}^{\mu}) +  O(\epsilon^4) \, , \nonumber\\
u_{I}^{a} =&  \bigg(1 +\frac{h^{(2)}_{tt}}{2}+ \frac{v_{I}^2}{2}\bigg)(1;v_{I}^{\mu}) +  O(\epsilon^4) \, , \label{4-vel}
\end{align} 
where $v_M$ and $v_I$ are the three-velocities of our two fluids, then we can write the components of the perturbed energy-momentum tensor as
 \begin{align}
T_{tt}=& \rho^{(2)} (1 - h^{(2)}_{tt}) + \rho^{(2)}_{I} v_{I}^2 + \rho^{(2)}_{M}v_{M}^2 \nonumber\\ 
& + \rho^{(2)}_{M} \Pi_{M}  +  \rho^{(4)}_{I} + p^{(2)}_{I} v_{I}^2 + O(\epsilon^6) \label{emtt} \, ,  \\  \nonumber\\ 
T_{t\mu} =& - \rho^{(2)}_{M} v_{M\mu} -( \rho^{(2)}_{I} + p^{(2)}_{I}) v_{I\mu} + O(\epsilon^5) \label{emtx} \, , \\   \nonumber\\ 
T_{\mu \nu} =& \rho^{(2)}_{M} v_{M\mu} v_{M\nu} + ( \rho^{(2)}_{I} + p^{(2)}_{I}) v_{I\mu} v_{I\nu} \nonumber \\ 
& +  (p^{(4)}_{M}  + p^{(2)}_{I} + p^{(4)}_{I}) g_{\mu\nu} + O(\epsilon^6) \, , \label{em}
\end{align}
where $\rho^{(2)} = \rho^{(2)}_{M} + \rho^{(2)}_{I}$, and where $\rho^{(2)}_{M}$ is the rest-mass energy density of the non-relativistic matter fields,  $\Pi_{M}$ is their specific energy density, and $p^{(4)}_{M}$ is their pressure. Similarly, ${\rho^{(2)}_{I}}$ and ${\rho^{(4)}_{I}}$ are the two lowest-order parts of the energy density of the barotropic fluid, and $p^{(2)}_{I}$ and $p^{(4)}_{I}$ are the two lowest-order contributions to its pressure. The reader may note the we have set $p^{(2)}_M=0$ for the non-relativistic matter fields, as we want this to represent dust-like sources such as galaxies and clusters.
 

Before considering Einstein's equations, we note that we can use the energy-momentum conservation equations for the non-interacting barotropic fluid to write
\begin{align}
\nabla p^{(2)}_{I} = 0 \, .\label{emcon1}
\end{align}
This is the leading-order part of the Euler equation of the barotropic fluid, and it immediately implies that both $p^{(2)}_{I}$ and $\rho^{(2)}_{I}$ must be functions of time only [as $p=p(\rho)$, for this fluid]. It also means that the leading-order part of the continuity equation for the barotropic fluid, which also follows directly from energy-momentum conservation, is given by
\begin{align}
\rho^{(2)}_{I,t} + (\rho^{(2)}_{I} + p^{(2)}_{I}) \nabla \cdot {\bm v}_{I} = 0 \, . \label{emcon2}
\end{align}
This is very similar to the conservation equation for a homogeneous fluid in FLRW models, and we later use it in the same way as that equation to determine the cosmological evolution.

\subsection{Einstein's field equations} \label{sec2a}

In order to find the geometry of the space-time within each cell, and to solve for the motion of its boundary, we need to use Einstein's field equations,
 \begin{align}
R_{ab} &= 8\pi G \left( T_{ab} - \frac{1}{2} T g_{ab} \right) + g_{ab}\Lambda \, , \label{Riccifield}
\end{align}
where $R_{ab}$ is the Ricci tensor, $g_{ab}$ is the metric of space-time, $\Lambda$ is the cosmological constant, $G$ is Newton's constant, $T_{ab}$ is the energy-momentum tensor, and $T= g^{ab} T_{ab}$ is its trace.

Using the perturbed metric given in Eq. \eqref{metric}, and the energy-momentum tensor from Eq. \eqref{emtt}, we can write the leading-order contributions to the $tt$-component of Einstein's equations as
\begin{align}
\nabla^2 h^{(2)}_{tt}  =  - 8\pi G\rho^{(2)} -24\pi G p^{(2)}_{I} + 2\Lambda \label{nablaphi} \, ,
\end{align}
where $\nabla^2 = \partial_{\alpha} \partial_{\alpha}$ is the three-dimensional Laplacian. Here we have taken the cosmological constant $\Lambda$ to contribute at $O(\epsilon^2)$, which means we are modelling a scenario where $\Lambda \sim  \rho^{(2)} \sim h^{(2)}_{tt}$. This happens on scales of about $100$ Mpc, where the cosmological constant is comparable to the background gravitational potential. This is still well below the cosmological horizon scale, where our post-Newtonian formalism is satisfied. 

The solution to Eq. \eqref{nablaphi} can be formally written as
\begin{align}
h^{(2)}_{tt} \equiv 2\Phi = 2\Phi_{M} + 2\Phi_{I} +6\Phi_{p}+ 2\Phi_{\Lambda} \, , \label{phi} 
\end{align}
where the potentials $\Phi_{M}$, $\Phi_{I}$, $\Phi_{p}$ and $\Phi_{\Lambda}$ are given implicitly as the solutions to
\begin{align}
\nabla^2 \Phi_{M} \equiv& - 4 \pi G \rho^{(2)}_{M} \, ,  \label{phim} \\
\nabla^2 \Phi_{I} \equiv& - 4 \pi G {\rho^{(2)}_{I}} \, ,  \label{phii} \\
\nabla^2 \Phi_{p} \equiv& - 4 \pi G p^{(2)}_{I} \, ,  \\
\nabla^2 \Phi_{\Lambda}  \equiv& \Lambda \, . \label{newpots}
\end{align}
Using the symmetries of our lattice model, and the fact that $p^{(2)}_{I}$ is a function of time only, the potentials $\Phi_{p_{I}}$ and $\Phi_{\Lambda}$ can be written explicitly as
\begin{align}
\Phi_{p_{I}}=& -\frac{2\pi G p^{(2)}_{I}}{3}(x^2 + y^2 + z^2) \ ,  \\
\Phi_{\Lambda}=& \frac{\Lambda}{6}(x^2 + y^2 + z^2) \, . \label{plamsolns}
\end{align}
Solutions to Eqs. (\ref{phim}) and (\ref{phii}) can be given in terms of Green's functions, as shown in \cite{vaas}. Auxiliary functions of time can also be added in $h_{tt}^{(2)}$, and absorbed into the matter potential, $\Phi_{M}$.

To go further, we now need to make a gauge choice. We make the following choice at $O(\epsilon^2)$, so that we remain as close as possible to the standard post-Newtonian gauge,
\begin{align} 
& \frac{1}{2} h^{(2)}_{tt,\mu} + h^{(2)}_{\mu\nu, \nu} - \frac{1}{2} h^{(2)}_{\nu \nu,\mu} = 3 \Phi_{p , \mu} +\frac{3}{2} \Phi_{\Lambda, \mu} \ .\label{gauge1}
\end{align}
This ensures that the metric is diagonal at $O(\epsilon^2)$, and there are no $O(\epsilon)$ contributions to the $t\mu$-component of the metric. Using Eqs. \eqref{em} and \eqref{plamsolns}, the $\mu\nu$-component of Einstein's equations can now be written as 
\begin{align}
\nabla^2 h^{(2)}_{\mu\nu} = - (8\pi G \rho^{(2)} +\Lambda)\delta_{\mu\nu} \, . \label{nablapsi}
\end{align}
The solution to this equation is given by
\begin{align}
h^{(2)}_{\mu\nu} \equiv 2 \Psi \delta_{\mu\nu} = (2\Phi_{M} + 2\Phi_{I} - \Phi_{\Lambda})\delta_{\mu\nu} \, .\label{psi} 
\end{align}
The reader may note that in this formalism we have $\Psi \neq \Phi$ in the presence of either a cosmological constant or a barotropic fluid (or both). This differs from the case of cosmological perturbation theory, where $\Phi = \Psi$ in the absence of anisotropic stress.

To solve for the ${t\mu}$-component of Einstein's equations, we now need to make a gauge choice at $O(\epsilon^3)$, which we do as follows:
\begin{align} 
& h^{(3)}_{\nu t, \nu} - \frac{1}{2} h^{(2)}_{\nu\nu,t} = 0  \, .  \label{gauge2}
\end{align}
Using both of our gauge conditions, Eqs. \eqref{gauge1} and \eqref{gauge2}, the $t\mu$-component of Einstein's equations can be written as
\begin{align}
& \nabla^2 h^{(3)}_{t\mu} +\Psi_{,t\mu}   =  16\pi G \left[ \rho^{(2)}_{M} v_{M\mu} + (\rho^{(2)}_{I} + p^{(2)}_{I}) v_{I\mu} \right]\, . \label{thirdorderpert}
\end{align}
The solution to this equation is given by
\begin{align}
h^{(3)}_{t\mu} = -4V_{M\mu} - 4V_{I\mu} + \frac{1}{2}\chi_{,t\mu} \, , \label{thirdsoln}
\end{align}
where we have used the two vector potentials
\begin{align}
\nabla^2 V_{M\mu} \equiv& -4\pi \rho^{(2)}_{M} v_{M\mu} \, , \\
\nabla^2 V_{I\mu} \equiv& - 4\pi G (\rho^{(2)}_{I} + p^{(2)}_{I}) v_{I\mu} \, ,
\end{align}
and the superpotential
\begin{align}
\nabla^2 \chi \equiv -2\Psi \, .
\end{align}
The gauge conditions imply that the divergence of these vector potentials must obey $V_{M\mu,\mu} + V_{I\mu, \mu} = - \Psi_{,t}$.

Finally, we can write the $O(\epsilon^4)$ part of the tt-component of Einstein's equations. Using the energy-momentum tensor from Eq. \eqref{em}, both our gauge conditions, and the lower-order solutions for $h_{tt}^{(2)}$, $h_{\mu \nu}^{(2)}$ and $h_{t\mu}^{(3)}$, this equation becomes
\begin{widetext}
\begin{align}
\nabla^2 h_{tt}^{(4)} = & -2\nabla(\Phi \nabla \Phi)  - \nabla (\Psi \nabla \Phi + \Phi \nabla \Psi) +4\pi G\rho^{(2)}\Phi +24\pi G p^{(2)}_{I}\Phi - \frac{5}{2}\Lambda\Phi -20\pi G\rho^{(2)}\Psi -60\pi G p^{(2)}_{I}\Psi   \label{nablasqhtt4}  \\
&\quad + 5\Lambda\Psi - 16\pi G\rho^{(2)}_{M}v_{M}^2  - 16\pi G\rho^{(2)}_{I}v_{I}^2 -8\pi G\rho^{(2)}_{M}\Pi_{M}   -8\pi G\rho^{(4)}_{I} -16 \pi G p^{(2)}_{I}v_{I}^2 -24\pi G p^{(4)}_{M} -24\pi G p^{(4)}_{I} \, . \nonumber
\end{align}
\end{widetext}
These equations can also be solved using the Green's functions from \cite{vaas}. Once this has been done, and the distribution of matter has been specified, this gives us sufficient information to find the geometry of each of our lattice cells, to post-Newtonian order of accuracy. 

Nowhere in this analysis have we assumed asymptotic flatness, as is conventionally done when applying the post-Newtonian formalism to the case of isolated systems. Instead, we have a system of equations that can be directly applied to solve for the gravitational fields of astrophysical bodies in a cosmological setting.

\section{Cosmological Expansion}  \label{sec3}

In this section, we derive the acceleration and constraint equations for the boundary of each of our cells, up to the first post-Newtonian level of accuracy. Due to the periodicity of our lattice models, these equations will also describe the large-scale expansion of the Universe as a whole. At Newtonian order, these equations take exactly the same form as the acceleration and constraint equations of a FLRW universe containing dust, a barotropic fluid, spatial curvature and a cosmological constant. At first post-Newtonian order, we obtain the leading-order corrections to these equations in a lattice universe.

Using reflection symmetric boundary conditions, as we do in this study, implies that the extrinsic curvature of each of the $(2+1)$-dimensional boundaries of every cell must vanish (see \cite{vaas} for details). This condition leads directly to the equation of motion of the cell boundary, which to post-Newtonian accuracy can be written as follows:
\begin{align} 
X_{,tt} =&  \bigg[ \Phi_{,x} - 2\Psi\Phi_{,x}+ \frac{h_{tt,x}^{(4)}}{2} - h_{tx,t}   -(2\Phi_{,x} + \Psi_{,x}) X_{,t}^{2} \nonumber \\
&\quad -( 2\Psi_{,t}+\Phi_{,t}) X_{,t}   - X^{(2)}_{,A} \Phi_{,A}\bigg] \bigg|_{x=X} + O(\epsilon^6) \, , \label{X1}  
\end{align}
which can also be derived from the geodesic equation. Likewise, we obtain a set of equations that describes the spatial curvature of the cell boundaries, and their rate of change, as
\begin{align}
X_{,AB} =& \delta_{AB} ( \Psi_{,x})|_{x=X} + O(\epsilon^4) \, , \label{X3} 
\end{align}
and
\begin{align}
X_{,tA} =& 
\frac{1}{2} \bigg[ h_{tA,x} -  h_{tx,A} - 2(\Phi_{,A} +\Psi_{,A}) X_{,t}\bigg]\bigg|_{x=X} + O(\epsilon^5) \, . \label{X2} 
\end{align}
Each of the quantities in these equations must be evaluated on the boundary of the cell. Together, they give us enough information to relate the evolution of the boundaries of our cells to the matter content within them. We will now do this to Newtonian, and then post-Newtonian, levels of accuracy.

\subsection{Newtonian accuracy}

For a regular polyhedron, at the Newtonian order of accuracy, the total surface area and volume of a cell are given by $A=\alpha_{\kappa}X^2$  and $V=\frac{1}{3} \alpha_{\kappa}X^3$, where $\alpha_{k}$ is a set of constants that depend on the cell shape in question (numerical values can be found in \cite{vaas}). By applying Gauss' theorem, and using Eq. \eqref{nablaphi}, we can re-write the evolution equation for $X$ as
\begin{align} 
X_{,tt} =  \frac{-4\pi G M - 4\pi G \int_{V}({\rho^{(2)}_{I}}+3p^{(2)}_{I}) \ dV^{(0)} }{\alpha_{\kappa}X^2} + \frac{\Lambda}{3} X \, , \label{acc1}
\end{align}
where $M$ is the gravitational mass of the non-relativistic matter, defined by $M\equiv \int_V \rho^{(2)}_{M}  \ dV^{(0)}$, the integrals are over the spatial volume interior to the cell, and $dV^{(0)}$ is the spatial volume element at zeroth order.
 
This equation can be simplified, and integrated, by making use of Reynold's transport theorem. This theorem states that for any function on space-time, $f$, we have
\begin{align}
\frac{d}{dt} \int f \ dV = \int f_{,t} \ dV + \int f \bm{v} \cdot d\bm{A} \, .
\end{align}
Taking $f$ to be the energy density, $\rho^{(2)}_{I}$, and using the conservation equations \eqref{emcon1} and \eqref{emcon2}, then gives
\begin{align}
\frac{d \int \rho^{(2)}_{I} \ dV}{dt} = - \int p^{(2)}_{I}  \bm{v_{I}} \cdot d \bm{A} = -p^{(2)}_{I}  X_{,t} A \, . \label{halfcont}
\end{align}
where we have required the barotropic fluid to be co-moving with the boundary of the cell, at all points on the boundary, and where we have made use of the fact that $\rho^{(2)}_I$ and $p^{(2)}_I$ are functions of time only. We then have the following conservation equation for the barotropic fluid
\begin{align}
\rho^{(2)}_{I,t}  + 3 \frac{X_{,t}}{X} ({\rho^{(2)}_{I}} + p^{(2)}_{I} )=0 \, .  \label{continuity1}
\end{align}
This is strongly reminiscent of the corresponding equation in FLRW cosmology, as it should be.
 
We can now simplify the evolution equation (\ref{acc1}), and integrate it using the continuity equation (\ref{continuity1}), to get
\begin{align} 
\frac{X_{,tt}}{X}&= \frac{-4\pi G M}{\alpha_{\kappa}X^3} - \frac{4\pi G}{3} ({\rho^{(2)}_{I}}+3p^{(2)}_{I}) +\frac{\Lambda}{3}\, , \label{acc}
\end{align}
and
\begin{align}
\left( \frac{X_{,t}}{X} \right)^2 &=   \frac{8\pi G M}{\alpha_{\kappa}X^3}  + \frac{8\pi G}{3} \rho^{(2)}_{I} - \frac{C}{X^2} + \frac{\Lambda}{3} \, , \label{constraint}
\end{align}
where $C$ is an integration constant. These equations are identical to the acceleration and constraint equations of an FLRW universe filled with dust, a barotropic fluid, and a cosmological constant, with $C$ taking the role of the spatial curvature.

Finally, using Eqs. \eqref{emcon2} and \eqref{continuity1}, we can read off that $\nabla . \bm{v}_{I} = 3  {X_{,t}}/{X}$. The three-velocity of the barotropic fluid is therefore given by
\begin{align}
\bm{v}_{I}^{\mu} =   \frac{X_{,t}}{X} (x, y, z) \, . \label{vel}
\end{align}
This expression will be very useful for evaluating some of the more complicated post-Newtonian expressions that will follow.

\subsection{Post-Newtonian accuracy}

In this section we calculate the post-Newtonian contributions to the equations of motion of the boundary, following a similar approach to the one used in \cite{vaas}. The principal difference in the present case is the inclusion of the barotropic fluid, and of $\Lambda$. These lead directly to extra terms in the energy-momentum tensor, but also result in $\Phi \neq \Psi$. We must therefore keep track of each of these potentials separately.
 
We begin by observing that the functional form of $X$, up to $O(\epsilon^2)$, is given by
\begin{align} 
X = \zeta + \frac{1}{2} (y^2+ z^2) \bm{n} \cdot \nabla \Psi + O(\epsilon^4) \, , \label{xfunc}
\end{align}  
where $\zeta=\zeta(t)$ is a function of time only, and corresponds to the position of the centre of a cell face in the $x$-direction. This observation follows from the lowest order parts of Eqs. \eqref{X1} - \eqref{X2}, from the gauge conditions \eqref{gauge1} and \eqref{gauge2}, and from symmetry arguments imposed at the centre of the cell face.

Taking time derivatives of Eq. \eqref{xfunc}, and substituting in from Eq. \eqref{X1}, then gives 
\begin{align} 
\zeta_{,tt} =& X_{,tt}  - \frac{1}{2}(y^2+ z^2)(\bm{n} \cdot \nabla  \Psi)^{\ddot{}}  + O(\epsilon^6) \ \nonumber
\\  =& \Phi_{,x} - 2\Psi\Phi_{,x} + \frac{h^{(4)}_{tt,x}}{2} - h_{tx,t}   -(2\Phi_{,x} + \Psi_{,x}) X_{,t}^{2}   \nonumber \\
& -( 2\Psi_{,t}+\Phi_{,t}) X_{,t} - X^{(2)}_{,A} \Phi_{,A} \nonumber \\
&- \frac{1}{2}(y^2+ z^2)(\bm{n} \cdot \nabla \Psi)^{\ddot{}} + O(\epsilon^6) \ ,  \label{zetatt}
\end{align}
where $^{.}$ represents a time derivative along the boundary and where all quantities in this equation should be evaluated on the boundary of the cell.

Several of the terms in Eq. \eqref{zetatt} can be related to the matter content within the cell by an application of Gauss' theorem. For example, we can use Eq. \eqref{nablapsi} to obtain
\begin{align} 
\bm{n} \cdot \nabla \Psi 
&= -\frac{4\pi G M}{\alpha_{\kappa}X^2} -\frac{4\pi G {\rho^{(2)}_{I}}  X}{3}  -\frac{\Lambda X}{6} \, . \label{normpsi} 
\end{align}
We can also replace a number of terms in Eq. \eqref{zetatt} using either the gauge condition, given in Eq. \eqref{gauge2}, or the lower-order solutions given in Eqs. \eqref{acc} and \eqref{constraint}. As an example of this, we can replace the $h_{tx,t}$ term in Eq. \eqref {zetatt} by using Eq. \eqref{gauge2} and Gauss' theorem. This gives
\begin{align}  
\kappa \int_{S} n_{\alpha}h_{ t\alpha, t} \ dS &=  \int_{\Omega} 3\Psi_{,tt} \ dV \, . \label{gauge3}
\end{align}
Finally, using the lower-order solutions for $\Phi$ and $\Psi$, from Eqs. \eqref{phi} and \eqref{psi}, we can write the generalized form of the acceleration equation in terms of the potentials defined in Eqs. \eqref{phim}-\eqref{newpots}. This gives
\begin{widetext}
\begin{align} 
X_{,tt} 
=&-\frac{4\pi G M}{A} +  (-4\pi G {\rho^{(2)}_{I}} -12\pi G p^{(2)}_{I}  + \Lambda )\frac{V}{A} - \frac{3\kappa}{\alpha_{\kappa}X^2}  \int_{S} \bigg( (\Phi_{M} + \Phi_{I} + \Phi_{p_{I}})_{,t} X_{,t}  \bigg) \ dS \nonumber \\
& +\frac{\kappa}{\alpha_{\kappa}X^2}  \int_{S} \bigg(2\Phi_{M} + 2\Phi_{I} + 3\Phi_{p_{I}} + \frac{1}{2}\Phi_{\Lambda}\bigg)\bigg(\frac{4\pi G M}{\alpha_{\kappa}X^2} + \frac{4\pi G}{3} ({\rho^{(2)}_{I}}+3p^{(2)}_{I} ) X -\frac{\Lambda X}{3}  \bigg) \ dS \nonumber \\
&+ \frac{1}{\alpha_{\kappa}X^2}\bigg[4\pi G \avg{{\rho^{(2)}_{I}}(\Phi_{M}+ \Phi_{I} + 3\Phi_{p_{I}}+ \Phi_{\Lambda})} +4\pi G \avg{\rho^{(2)}_{M} (-2\Phi_{M}-2\Phi_{I} + 3\Phi_{p_{I}}+ \frac{5}{2} \Phi_{\Lambda})} \nonumber \\
&\qquad \qquad + 12\pi G \avg{p^{(2)}_{I} (\Phi_{M}+ \Phi_{I} + 3\Phi_{p_{I}}+ \Phi_{\Lambda})} - \avg{\Lambda (\Phi_{M}+ \Phi_{I} + 3\Phi_{p_{I}}+ \Phi_{\Lambda})}  -12\pi G \avg{p^{(4)}_{M}} \nonumber \\
& \qquad \qquad - 8\pi G\avg{\rho^{(2)}_{M}v_{M}^2}  - 8\pi G\avg{\rho^{(2)}_{I}v_{I}^2} -4\pi G\avg{\rho^{(2)}_{M}\Pi_{M}}  -4\pi G\avg{\rho^{(4)}_{I}} -8 \pi G \avg{p^{(2)}_{I}v_{I}^2}    -12\pi G\avg{ p^{(4)}_{I}} \bigg]\nonumber \\
& + \frac{96\pi^2 G^2 M^2}{\alpha_{\kappa}^2 X^3} + \frac{64\pi^2 G^2 M {\rho^{(2)}_{I}}}{\alpha_{\kappa}} - \frac{12\pi G M C}{\alpha_{\kappa}X^2} + \frac{32\pi^2 G^2 {\rho^{(2)}_{I}}^2 X^3}{3} - 4\pi G {\rho^{(2)}_{I}} C X \nonumber \\
&+ \frac{64\pi^2 G^2 M p^{(2)}_{I} }{\alpha_{\kappa}} + \frac{8}{3}\pi G p^{(2)}_{I} \Lambda X^3 + \frac{64\pi^2 G^2 {\rho^{(2)}_{I}} p^{(2)}_{I} X^3}{3} - 8\pi G p^{(2)}_{I} C X -\frac{\Lambda^2 X^3}{6}  + \frac{\Lambda C X }{2} \nonumber \\ 
& - \frac{3}{\alpha_{\kappa}X^2}\int_{V} (\Phi_{M} + \Phi_{I} - \frac{1}{2}\Phi_{\Lambda})_{,tt} \ dV - \frac{1}{2} (\bm{n} \cdot \nabla \Psi)^{\ddot{}} \bigg[\frac{\kappa}{\alpha_{\kappa}X^2} \int_{S} (y^2+ z^2) \ dS - (y^2 + z^2)\bigg] + O(\epsilon^6) \ , \label{final_pots}
\end{align}
where $V$ is the volume of the cell, $A$ is the total surface area of the cell, and $\kappa$ is the number of faces of the cell. The notation $\avg{\varphi} =  \int_V \varphi \ dV$ is used to denote quantities integrated over the volume interior to the cell, where $\varphi$ is some scalar function on the space-time. The quantity $( \bm{n} \cdot \nabla \Psi)^{\ddot{}}$, in this equation, can be found to be given by
\begin{align}
( \bm{n} \cdot \nabla \Psi)^{\ddot{}}&=  -\frac{224\pi^2 G^2 M^2}{\alpha_{\kappa}^2 X^5} -\frac{14\pi G M \Lambda}{3\alpha_{\kappa} X^2}  -\frac{448\pi^2 G^2 M {\rho^{(2)}_{I}}}{3\alpha_{\kappa}X^2}+\frac{24\pi G MC}{\alpha_{\kappa}X^4} -\frac{112\pi^2 G^2 M p^{(2)}_{I} }{\alpha_{\kappa} X^2} \nonumber \\ 
  &\quad   - \frac{224\pi^2 G^2 {\rho^{(2)}_{I}}^2 X}{9} - \frac{112\pi^2 G^2 {\rho^{(2)}_{I}} p^{(2)}_{I} X}{3} - \frac{14 \pi G {\rho^{(2)}_{I}} \Lambda X}{9} - 16\pi^2 G^2 {p^{(2)}_{I}}^2 X \nonumber \\
  &\quad -\frac{2\pi G p^{(2)}_{I} \Lambda X }{3} - \frac{\Lambda^2  X}{18} + \frac{8 \pi G {\rho^{(2)}_{I}} C}{X} + \frac{8 \pi G p^{(2)}_{I} C}{X} +  4\pi G p^{(2)}_{I, t} X_{,t} \ .     \label{psiddot}
\end{align}
The acceleration equation \eqref{final_pots} is fully general, being valid for any cell shape and any distribution of matter in the presence of a barotropic fluid and a cosmological constant. This complicated equation reduces to the one derived in \cite{vaas}, in the absence of the barotropic fluid and the cosmological constant. In addition, however, the present equation contains several cross terms between the different types of matter. These arise due to the non-linearity of Einstein's equations, and should be expected to alter the effects of back-reaction.

Before moving on to consider simple matter distributions, we can simplify Eq. \eqref{final_pots} a little by looking at the specific case of cubic cells. In this case the total volume of a cell is given by
\begin{align}
V = 8 \zeta^3 + 8 (\bm{n} \cdot \nabla \Psi) \zeta^4 + 3\int_{V} \Psi \ dV + O(\epsilon^4) \, ,
\end{align}
and the total surface area is given by
\begin{align}
A&= 24\zeta^2\bigg(1 + \frac{4}{3} (\bm{n} \cdot \nabla \Psi) \zeta + \frac{1}{2\zeta^2} \int_{S} \Psi \ dS\bigg) + O(\epsilon^4) \, .
\end{align}
We can also use $\kappa = 6$ and $\alpha_{k}=24$, for the specific case of cubic cells, and rewrite the acceleration equation \eqref{final_pots} as
\begin{align}
X_{,tt} =&\frac{-  \pi G M}{6 \zeta^2} -\frac{4\pi G}{3}( {\rho^{(2)}_{I}} + 3 p^{(2)}_{I}) \zeta + \frac{\Lambda\zeta}{3} \nonumber \\ 
&+ \frac{7\pi^2 G^2 M^2}{27 X^3} +\frac{118\pi^2 G^2 M{\rho^{(2)}_{I}} }{27}  +\frac{5\pi G M\Lambda}{108} + 4\pi^2 G^2 M p^{(2)}_{I}   + \frac{496\pi^2 G^2 {\rho^{(2)}_{I}}^2 X^3}{27} + 32\pi^2 G^2 {\rho^{(2)}_{I}}P X^3\nonumber \\
&+\frac{16\pi G {\rho^{(2)}_{I}} \Lambda X^3}{27} + \frac{8\pi G p^{(2)}_{I} \Lambda X^3}{3} - \frac{7\Lambda^2 X^3}{54} - \frac{5\pi G M C}{6X^2} - \frac{20 \pi G {\rho^{(2)}_{I}} C X}{3} - \frac{32 \pi G p^{(2)}_{I} C X}{3} +\frac{\Lambda C X }{2} \nonumber \\
& +\frac{1}{4X^2}  \int_{S} \bigg(4\Phi_{M} + 4\Phi_{I} + 3\Phi_{p_{I}} - \frac{1}{2}\Phi_{\Lambda}\bigg)\bigg(\frac{\pi G M}{6X^2} + \frac{4\pi G}{3} ({\rho^{(2)}_{I}}+3p^{(2)}_{I} ) X - \frac{\Lambda X}{3}  \bigg) \ dS \nonumber \\
&-\frac{3}{4X^2}  \int_{S} \bigg((\Phi_{M} + \Phi_{I} + \Phi_{p_{I}})_{,t} X_{,t}  \bigg)  \ dS  +\frac{16\pi^2 G^2 {p^{(2)}_{I}}^2 X}{3} - \frac{ 4\pi G p^{(2)}_{I, t} X_{,t} X^2}{3} \nonumber \\
&+ \frac{1}{24X^2}\bigg[\avg{(4 \pi G (\rho^{(2)} + 3p^{(2)}_{I} ) - \Lambda)(- 2\Phi_{M} - 2\Phi_{I} + 3\Phi_{p_{I}}+ \frac{5}{2}\Phi_{\Lambda})} - 8\pi G\avg{\rho^{(2)}_{M}v_{M}^2}  - 8\pi G\avg{\rho^{(2)}_{I}v_{I}^2}   \nonumber \\
&  \qquad \qquad -4\pi G\avg{\rho^{(2)}_{M}\Pi_{M}}  -4\pi G\avg{\rho^{(4)}_{I}} -8 \pi G \avg{p^{(2)}_{I}v_{I}^2}-12\pi G \avg{p^{(4)}_{M}} -12\pi G\avg{ p^{(4)}_{I}} \bigg]\nonumber \\
 & - \frac{1}{8X^2} \int_{V} (\Phi_{M} + \Phi_{I} - \frac{1}{2}\Phi_{\Lambda})_{,tt} \ dV + \frac{1}{2} (\bm{n} \cdot \nabla \Psi)^{\ddot{}} (y^2 + z^2) + O(\epsilon^6) \ . \label{cubicacc}
\end{align}
Every term in this equation can be solved for in complete generality using the Green's function formalism set out in \cite{vaas}, but it still remains a very complicated expression. Instead, and in order to show the effects of back-reaction in a simple illustrative example, we look at the case of regularly arranged point-like particles in a sea of radiation, and in the presence of a cosmological constant.

\section{Point sources with radiation, spatial curvature and $\Lambda$} \label{sec4}

To find an explicit solution to the acceleration equation, let us consider the case of a point source located at the centre of each cell, in the presence of radiation and a cosmological constant. To simplify matters further, let us evaluate the acceleration equation at the centre of a cell face ({\it i.e.} at $y=z=0$). 

\subsection{Solutions}
In the case of point sources we have $v_{M}^{\alpha} = p^{(4)}_{M} = \Pi_{M}  =\avg{\rho^{(2)}_{M}  \Phi_{M}} =\avg{\rho^{(2)}_{M}  \Phi_{I}}= \avg{\rho^{(2)}_{M}  \Phi_{p_{I}}} = \avg{\rho^{(2)}_{M}  \Phi_{\Lambda}} = 0$. Hence, in this case, the potentials defined in Eq. \eqref{newpots} simplify to
\begin{align} \label{110}
\nabla^2 \Phi_{M} = -4\pi G M \delta(\mathbf{x}) \, , \quad
\nabla^2 \Phi_{I} = -4\pi G {\rho^{(2)}_{r}} \, , \quad 
\nabla^2 \Phi_{p_{I}} = -\frac{4\pi G}{3} \rho^{(2)}_{r} \, , \quad {\rm and} \quad 
\nabla^2 \Phi_{\Lambda} = \Lambda \, ,
\end{align}
where $M$ is the gravitational mass of the point source at the centre of the cell, and ${\rho^{(2)}_{r}}$ is the energy density of the radiation. The first of these potentials can be solved for, using the method of images, and can be used to absorb all auxiliary functions of time (see \cite{vaas} for details). This gives
\begin{align}  
\Phi_{M}  &=    \lim _{\mathcal{N} \to \infty} \sum_{\bm{\beta} = - \mathcal{N}}^{\mathcal{N}}\frac{G M}{\sqrt{(x -2\beta_{1} X)^2 + (\hat{y} -2 \beta_{2} X)^2 + (\hat{z}- 2\beta_{3} X)^2 }}  -   \lim _{\mathcal{N} \to \infty} \sum_{\bm{\beta^{*}} = - \mathcal{N}}^{\mathcal{N}}\frac{G M}{2 |\bm{\beta}| X} \, , \label{intphi}
\end{align}
where  $\bm{\beta^{*}}$ indicates that the null triplet has been removed. The remaining potentials are given by
\begin{align}
\Phi_{I}=  -\frac{2\pi G {\rho^{(2)}_{r}}}{3}(x^2 + y^2 + z^2)  \, , \quad
\Phi_{p_{I}}= -\frac{2\pi G \rho^{(2)}_{r}}{9}(x^2 + y^2 + z^2) \, , \quad {\rm and} \quad
\Phi_{\Lambda}= \frac{\Lambda}{6}(x^2 + y^2 + z^2) \, . \label{pot_solns}
\end{align}

If we now assume that the radiation does not interact with the point sources, then we have $p^{(4)}_{r} = \frac{1}{3}\rho^{(4)}_{r}$. Using the energy-momentum conservation equation at $O(\epsilon^4)$, the velocity of the barotropic fluid given in Eq. \eqref{vel}, and the lower-order acceleration and constraint equations, the energy density of radiation at $O(\epsilon^4)$ can then be seen to be given by
\begin{align}
\rho^{(4)}_{r} 
=& \bigg[\frac{\pi G M \rho^{(2)}_{r}}{X^3} + \frac{16}{3}\pi G {\rho^{(2)}_{r}}^2  - 2 \frac{\rho^{(2)}_{r}C}{X^2} +  \frac{2\rho^{(2)}_{r}\Lambda}{3}  \bigg] r^2 + 4\rho^{(2)}_{r}\Phi_{M} \, , \label{rho4}
\end{align}
where $r^2=x^2 + y^2 + z^2$. Using all of this information, the acceleration equation \eqref{cubicacc} can then be found to reduce to
\begin{align}
X_{,tt}=&-\frac{\pi G M}{6X^2} - \frac{8\pi G}{3} \rho^{(2)}_{r} X +\frac{\Lambda X}{3} 
+\frac{\pi G^2 M^2}{X^3} \mathcal{A}_{1}  + \pi G^2 M\rho^{(2)}_{r} \mathcal{A}_{2} +  G M \Lambda \mathcal{A}_{3} \nonumber \\
& + \frac{G M C}{X^2} \mathcal{A}_{4} - \frac{64}{9}\pi^2 G^2 {\rho^{(2)}_{r}}^2 X^3 - \frac{8}{9}\pi G \rho^{(2)}_{r} \Lambda X^3 -\frac{2}{9} \Lambda^2 X^3 +\frac{4}{3} \pi G \rho^{(2)}_{r} C X  + \frac{1}{2} \Lambda C X + O(\epsilon^6) \ , \label{accmink}
\end{align}
where $\mathcal{A}_{1}$, $\mathcal{A}_{2}$, $\mathcal{A}_{3}$ and $\mathcal{A}_{4}$ are constants whose values are given in Table \ref{tab1}, and whose relationship to the variables used in \cite{vaas} are given in the appendix. 
\end{widetext}

\begin{table}[b!]
\begin{tabular}{ | c | l |  }
    \hline 
    \textbf{\, Constant \,} & \textbf{\, Numerical value \,} \\ \hline 
    $\mathcal{A}_{1}$ & $\qquad \phantom{-}1.27 \ldots$   \\ \hline 
    $\mathcal{A}_{2}$ & $\qquad  -9.29 \dots$ \\  \hline
    $\mathcal{A}_{3}$ & $\qquad -0.219 \dots$  \\ \hline
    $\mathcal{A}_{4}$ & $\qquad \phantom{-}0.809 \dots$ \\ \hline
\end{tabular}
  \caption{\label{tab1} The numerical values of $\mathcal{A}_{1}$, $\mathcal{A}_{2}$, $\mathcal{A}_{3}$ and $\mathcal{A}_{4}$, from Eq. \eqref{accmink}. These are the numbers approached as the number of reflections in the method of images diverges to infinity.} 
\end{table}

Although already dramatically simplified, we can reduce this equation further by transforming into a FLRW background. This can be achieved using the following coordinate transformations \cite{vaas}:
\begin{align}
t &= \hat{t} + \frac{a_{,\hat{t}} a}{2} (\hat{x}^2 + \hat{y}^2 + \hat{z}^2) + O(\epsilon^3)  \label{timetrans} \\ \nonumber \\ 
x &= a \hat{x} \bigg[1 + \frac{(a_{,\hat{t}})^2}{4} (\hat{x}^2 + \hat{y}^2 + \hat{z}^2)\bigg] +O(\epsilon^4)   \\ \nonumber \\
y &= a \hat{y} \bigg[1 + \frac{(a_{,\hat{t}})^2}{4} (\hat{x}^2 + \hat{y}^2 + \hat{z}^2)\bigg] +O(\epsilon^4)  \\ \nonumber \\
z &= a \hat{z} \bigg[1 + \frac{(a_{,\hat{t}})^2}{4} (\hat{x}^2 + \hat{y}^2 + \hat{z}^2)\bigg] +O(\epsilon^4)  \, ,  
\end{align}
where the new coordinates $\hat{t}, \hat{x}, \hat{y}, \hat{z}$ are the standard set in an FLRW background, and where $a(\hat{t})$ is the scale factor of that background.

The energy density in these new coordinates is given by
\begin{align}
\rho^{(2)}_{r}(t) 
&= \rho^{(2)}_{r}(\hat{t}) - 2a_{,\hat{t}}^2 \rho^{(2)}_{r}(\hat{t}) (\hat{x}^2 + \hat{y}^2 + \hat{z}^2) +O(\epsilon^4)\, .
\end{align}
Evaluating this expression at the centre of a cell face, and using the the lower-order constraint equation \eqref{constraint}, gives
\begin{align}
\rho^{(2)}_{r}(t)  =& \hat{\rho}^{(2)}_{r} - \bigg(\frac{2\pi G (\hat{\rho}^{(2)}_{r}) M }{3 a \hat{X}_{0}^3} + \frac{16\pi G (\hat{\rho}^{(2)}_{r})^2 a^2}{3}  \label{rhotrans} \\
& \qquad \qquad \qquad \qquad + \frac{2\hat{\rho}^{(2)}_{r}\Lambda a^2}{3} - 2\hat{\rho}^{(2)}_{r}k\bigg) \hat{X}_{0}^2 \, . \nonumber
\end{align}
In this last equation we have introduced the abbreviated notation $\hat{\rho}^{(2)}_{r} = \rho^{(2)}_{r}(\hat{t})$, and used $k$ to denote Gaussian curvature in the background FLRW geometry.

%

Similarly, at the centre of a cell face the position of the boundary transforms as 
\begin{align} 
X=& a \hat{X}_{0}\bigg[1+ \frac{a_{,\hat{t}}^2}{4} \hat{X}_{0}^2\bigg] \label{XtoXhat} \\
=& a \hat{X}_{0} \bigg[1+ \bigg( \frac{\pi G M }{12 a \hat{X}_{0}^3} + \frac{2\pi G \hat{\rho}^{(2)}_{r} a^2}{3}  
+ \frac{\Lambda}{12} a^2 - \frac{k}{4} \bigg) \hat{X}_{0}^2 \bigg] ,  \nonumber
\end{align}
where $a=a(\hat{t})$ in this expression. 

\subsection{Results} \label{results}
Finally, using Eqs. \eqref{timetrans} - \eqref{XtoXhat}, the acceleration equation \eqref{accmink} simplifies down to
 \begin{align}
\frac{\ddot{a}}{a} = &-\frac{4\pi G}{3} (\hat{\rho}^{(2)}_{M} +2 \hat{\rho}^{(2)}_{r}) + \frac{\Lambda}{3} +\mathcal{B}_{1} + O(\epsilon^6) \, , \label{backacc}
\end{align}
where overdots in this equation denote derivatives with respect to $\hat{t}$, and where the back-reaction term, $\mathcal{B}_{1}$, is given by 
\begin{align}
\mathcal{B}_{1} 
&\simeq \left(4 \pi G \hat{\rho}^{(2)}_{M} a \hat{X}_{0}\right)^2 \left(1.50 -  1.20 \frac{\Omega_{r}}{\Omega_{M}}   + 0.88 \frac{\Omega_{k}}{\Omega_{M}}\right)  \, . \label{b1}
\end{align}
In writing these equations we have used the expression $\hat{\rho}^{(2)}_{M} \equiv {M}/{8 a^3 \hat{X}_{0}^3}$ for the average mass density in a cell, and have introduced the usual cosmological parameters
\begin{align}
\Omega_{M}  \equiv& \frac{8\pi G \hat{\rho}^{(2)}_{M}}{3 H^2} \ , \quad \Omega_{r}  \equiv \frac{8 \pi G \hat{\rho}^{(2)}_{r}}{3 H^2} \ , \quad \Omega_{k}  \equiv& -\frac{k}{a^2 H^2} \, , \nonumber
\end{align}
where $H \equiv \dot{a}/a$. The numerical values inside the brackets in Eq. \eqref{b1} are calculated from the constants in Table \ref{tab1}, and are quoted to the second decimal place only. The reader will note that $\Lambda$ does not appear in this expression, and so does not contribute to this back-reaction term at this level of accuracy. It can also be seen that, in the absence of the point-like particles, the acceleration equation reduces to the standard Friedmann equation for a universe with radiation, spatial curvature and a cosmological constant, as expected.


\begin{figure}[t!]
\includegraphics[width=0.52 \textwidth]{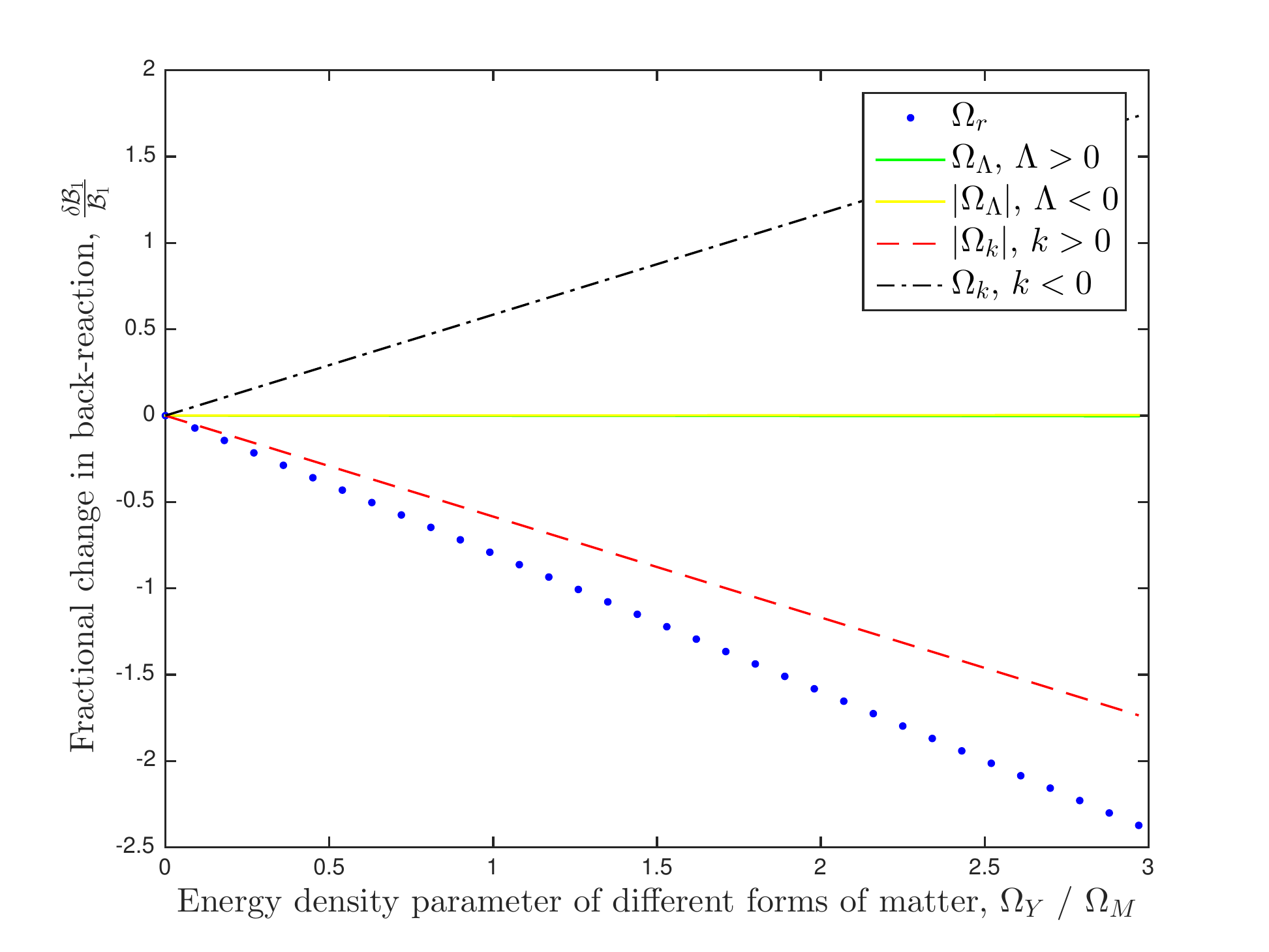}
\caption{\label{fig2} The effect of different forms of matter on the back-reaction term that appears in the acceleration equation, \eqref{backacc}. This is expressed in terms of the fractional change in $\mathcal{B}_{1}$. The energy density parameter $\Omega_{Y}$ for each type of matter is expressed as a fraction of $\Omega_{M}$.}
\end{figure}

\begin{figure}[b!]
\includegraphics[width=0.52 \textwidth]{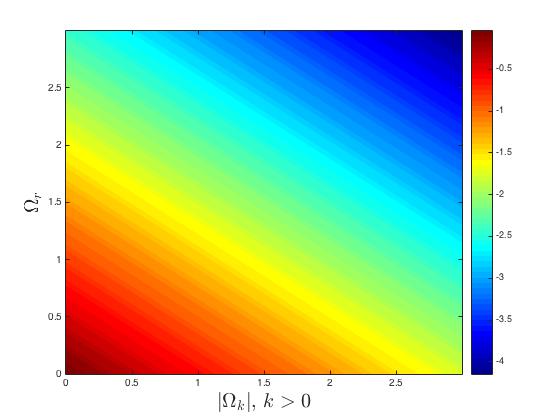}
\caption{\label{fig3} The effect that simultaneously adding radiation and positive spatial curvature has on the back-reaction term in the acceleration equation, ${\mathcal{B}_{1}}$.}
\end{figure}

\begin{figure}[b!]
\includegraphics[width=0.52 \textwidth]{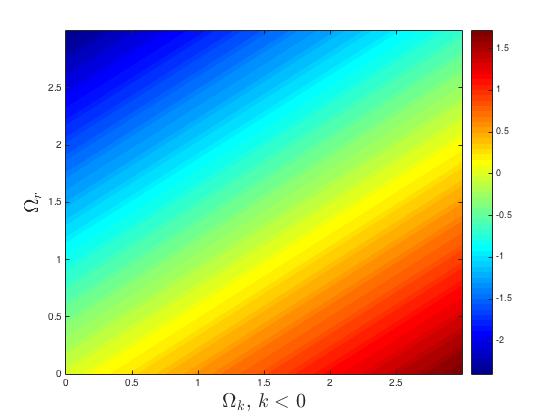}
\caption{\label{fig4} The effect that simultaneously adding radiation and negative spatial curvature has on the back-reaction term in the acceleration equation, ${\mathcal{B}_{1}}$.}
\end{figure}

The back-reaction term, $\mathcal{B}_{1}$,  is strongly influenced by the presence of radiation and spatial curvature, but not $\Lambda$. As can be seen from Fig. \ref{fig2}, the magnitude of $\mathcal{B}_{1}$ decreases as the amount of radiation in the Universe increases. This is independent of the expected suppression in the growth of structure that radiation is known to cause, as the discrete nature of the non-relativistic matter in this example exists for all time. Fig. \ref{fig2} also shows us that the back-reaction effect reduces for a closed universe, and increases for an open universe. In Fig. \ref{fig3} we plot the consequences of having non-zero amounts of both radiation and positive spatial curvature, while in Fig. \ref{fig4} we show the corresponding plot for negative  spatial curvature. In this latter case the spatial curvature and radiation can have compensating effects as they are simultaneously increased.

As well as an acceleration equation, we can integrate Eq. \eqref{backacc} to obtain a constraint equation. This is given by
 \begin{align}
\left( \frac{\dot{a}}{a} \right)^2 =& \frac{8\pi G}{3}( \hat{\rho}^{(2)}_{M} + \hat{\rho}^{(2)}_{r}) -\frac{k}{a^2} +\frac{\Lambda}{3} +\mathcal{B}_{2} + O(\epsilon^6) \, , \label{backcon}
\end{align}
where we have introduced $\mathcal{B}_{2}$ to denote the leading-order contribution to the back-reaction in this equation, and written $C=k\hat{X}_{0}^2 + O(\epsilon^4)$. The back-reaction term can be written explicitly as
\begin{align} \label{b2}
\mathcal{B}_{2} 
&\simeq -\left( 4\pi G \hat{\rho}^{(2)}_{M} a \hat{X}_{0}\right)^2 \left(1.50 -  0.80 \frac{\Omega_{r}}{\Omega_{M}} + 1.76 \frac{\Omega_{k}}{\Omega_{M}} \right)  \, .
\end{align}

\begin{figure}[t!]
\includegraphics[width=0.52 \textwidth]{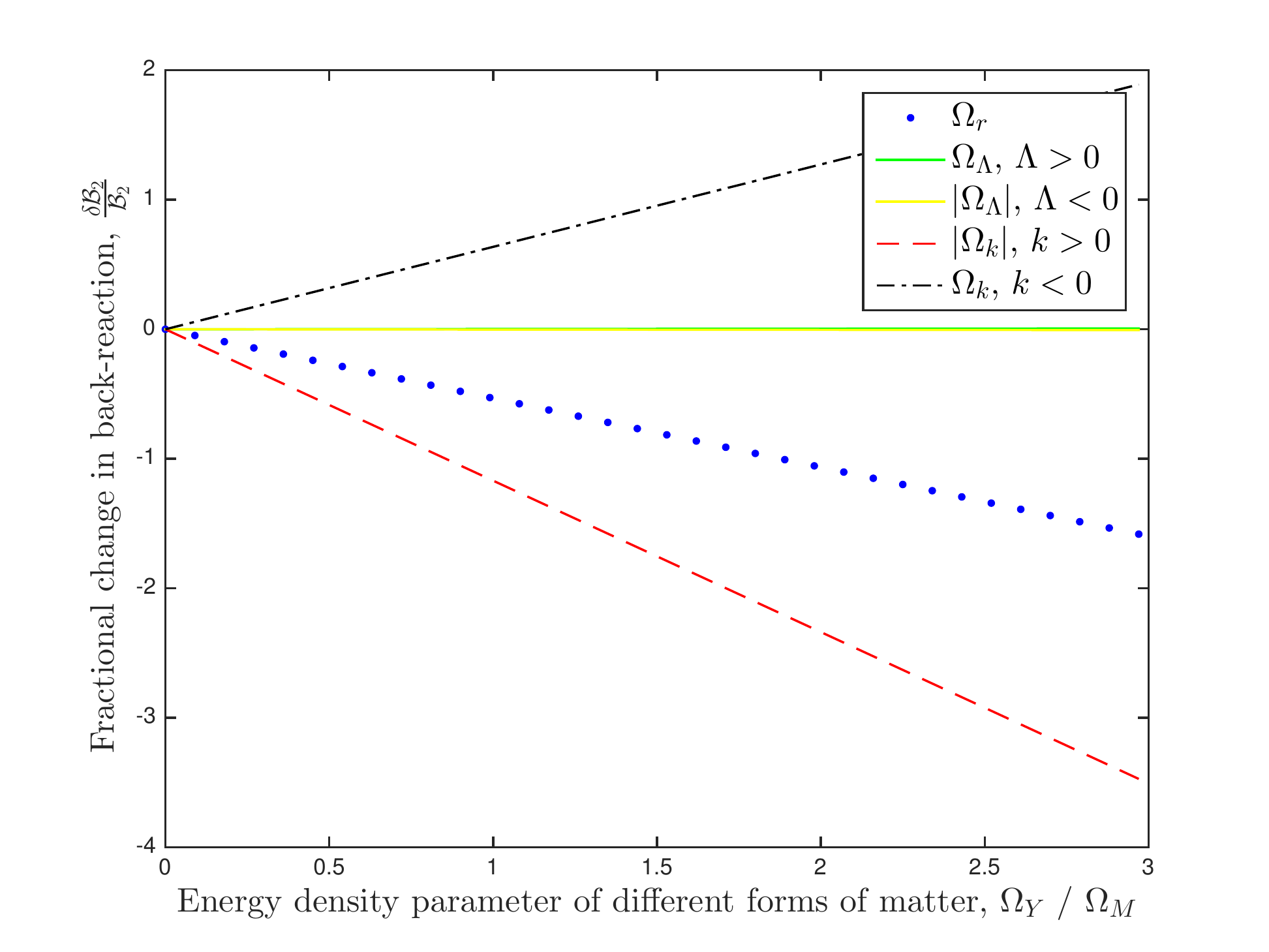}
\caption{\label{fig5} The effect of different forms of matter on the back-reaction term that appears in the constraint equation, \eqref{backcon}. This is expressed in terms of the fractional change in $\mathcal{B}_{2}$.}
\end{figure}

\begin{figure}[b!]
\includegraphics[width=0.52 \textwidth]{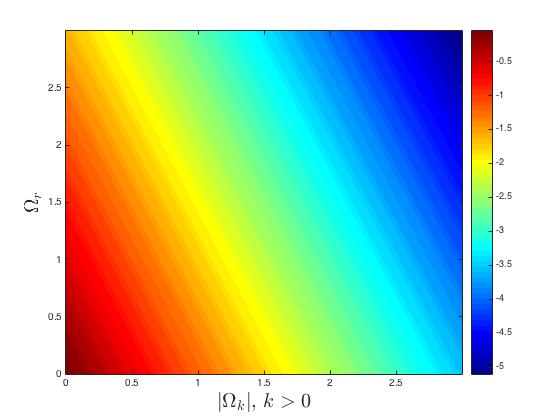}
\caption{\label{fig6} The effect that simultaneously adding radiation and positive spatial curvature has on the back-reaction term in the constraint equation, ${\mathcal{B}_{2}}$.}
\end{figure}

\begin{figure}[b!]
\includegraphics[width=0.52 \textwidth]{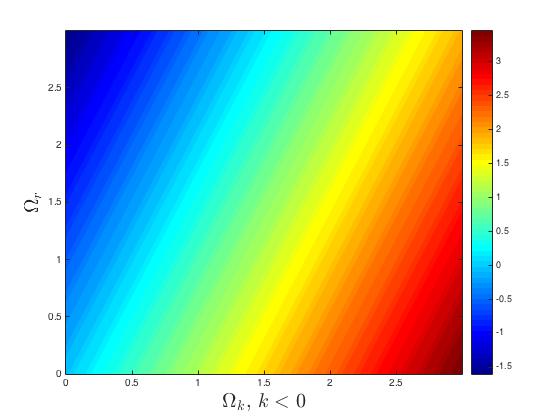}
\caption{\label{fig7} The effect that simultaneously adding radiation and negative spatial curvature has on the back-reaction term in the constraint equation, ${\mathcal{B}_{2}}$.}
\end{figure}

Let us now consider how different forms of matter affect the back-reaction in the Hubble rate. From Fig. \ref{fig5} it can be seen that the effect of radiation is to decrease the back-reaction term in this equation. In the Hubble rate, the back-reaction effect from the non-relativistic matter itself is negative. This means that radiation increases the value of the Hubble rate. The cosmological constant again makes a negligible contribution to the back-reaction. Finally, at $O(\epsilon^4)$, the Hubble rate is greater for a universe with positive spatial curvature, and smaller for a universe with negative spatial curvature. In Figs. \ref{fig6} and \ref{fig7} we plot the results of simultaneously adding radiation and spatial curvature. Once again, if spatial curvature is negative, then the effect it has on the back-reaction term can compensate that of radiation. If spatial curvature is positive, however, the effect it has on back-reaction is complementary to that of radiation.

Let us now consider the functional form of the different terms in the back-reaction equations. Recall that the lowest-order parts of the matter density and radiation density both scale in exactly the same way as in a FLRW model. This means that the leading-order correction arising from the non-relativistic matter itself is a radiation-like term, as identified in \cite{vaas}. The non-linear effect from radiation, on the other hand, scales as a fluid with equation of state $p= \frac{2}{3} \rho$. This is somewhere between the behaviour expected from a free scalar field, and that of normal radiation. The leading-order correction from the spatial curvature scales in the same way as non-relativistic matter, and effectively renormalises the value of the gravitational mass in the Universe.

\begin{figure}[t!]
\includegraphics[width=0.52 \textwidth]{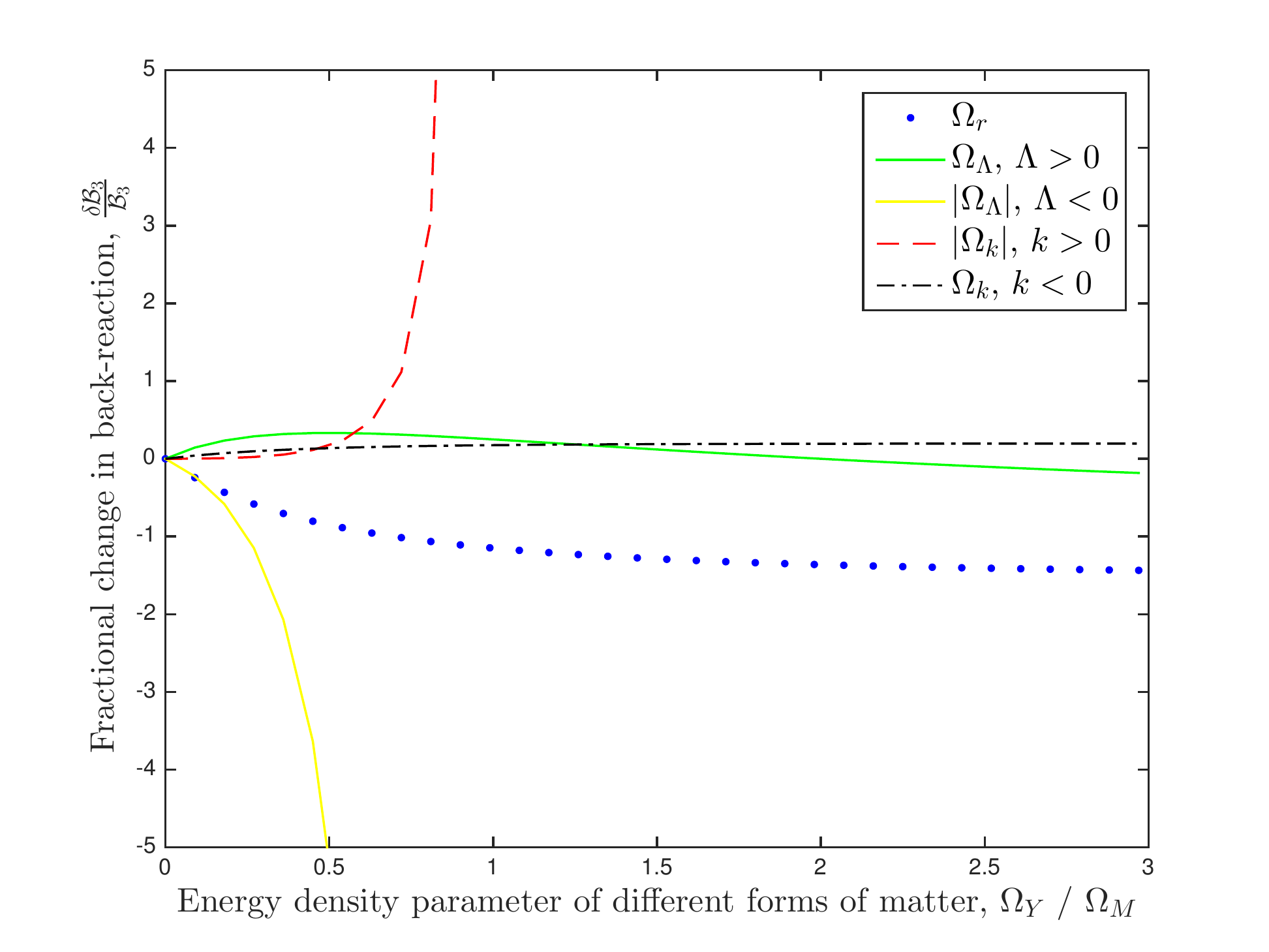}
\caption{\label{fig8} The effect of different forms of matter on the back-reaction term that appears in the deceleration parameter, \eqref{backdec}. This is expressed in terms of the fractional change in $\mathcal{B}_{3}$.}
\end{figure}


\begin{figure}[b!]
\includegraphics[width=0.52 \textwidth]{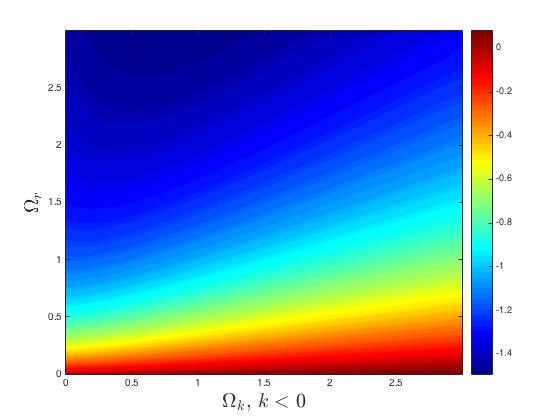}
\caption{\label{fig10} The effect that simultaneously adding radiation and positive spatial curvature has on the back-reaction term in the deceleration equation, ${\mathcal{B}_{3}}$.}
\end{figure}


\begin{figure}[t!]
\includegraphics[width=0.52 \textwidth]{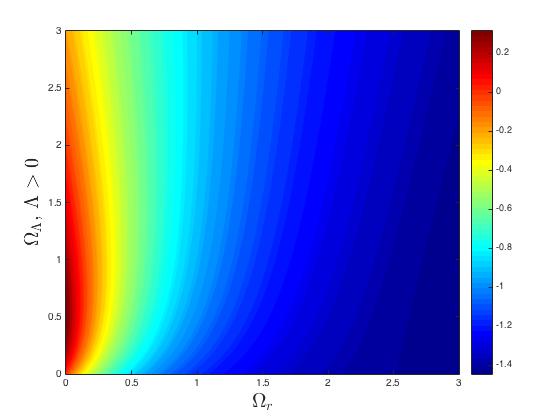}
\caption{\label{fig11} The effect that simultaneously adding radiation and a cosmological constant has on the back-reaction term in the deceleration equation, ${\mathcal{B}_{3}}$.}
\end{figure}

\begin{figure}[b!]
\includegraphics[width=0.52 \textwidth]{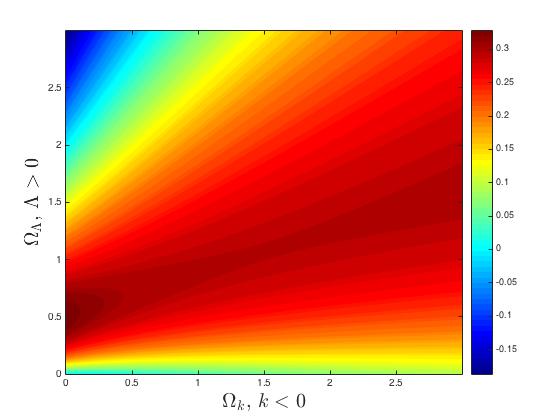}
\caption{\label{fig13} The effect that simultaneously adding radiation and negative spatial curvature has on the back-reaction term in the deceleration equation, ${\mathcal{B}_{3}}$.}
\end{figure}

Let us now consider the deceleration parameter, $q_0$. Using Eqs. \eqref{backacc} and \eqref{backcon}, we find this parameter to be given by
\begin{align}
q_{0} &\equiv - \frac{\ddot{a} a}{\dot{a}^2} 
= \frac{(\Omega_{M} +2 \Omega_{r} - 2 \Omega_{\Lambda} )}{ 2( \Omega_{M} + \Omega_{r} +  \Omega_{\Lambda} +  \Omega_{k})} + \mathcal{B}_{3} + O(\epsilon^4) \label{backdec} \, ,
\end{align}
where the back-reaction term in this equation is
\begin{align} \label{b3}
\mathcal{B}_{3} =& -\frac{3\mathcal{B}_{1}}{8\pi G \hat{\rho}^{(2)}_{M} \left( 1 + \frac{\Omega_{r}}{\Omega_{M}} +  \frac{\Omega_{\Lambda}}{\Omega_{M}} +  \frac{\Omega_{k}}{\Omega_{M}} \right) } \nonumber \\ \nonumber \\
&- \frac{3\mathcal{B}_{2} (1 +2 \frac{\Omega_{r}}{\Omega_{M}} - 2 \frac{\Omega_{\Lambda}}{\Omega_{M}} )}{16\pi G  \hat{\rho}^{(2)}_{M} \left( 1 + \frac{\Omega_{r}}{\Omega_{M} } +  \frac{\Omega_{\Lambda}}{\Omega_{M} } +  \frac{\Omega_{k}}{\Omega_{M}} \right)^2 } \, ,
\end{align}
where $\Omega_{\Lambda} \equiv \Lambda/3 H^2$, and where the values of $\mathcal{B}_{1}$ and $\mathcal{B}_{2}$ are given in Eqs. (\ref{b1}) and (\ref{b2}).

The effect that radiation, spatial curvature and a cosmological constant have on the back-reaction term $\mathcal{B}_{3}$ is displayed graphically in Fig. \ref{fig8}. Unlike the cases of $\mathcal{B}_{1}$ and $\mathcal{B}_{2}$, it can be seen that $\mathcal{B}_{3}$ is only of order $\epsilon^2$. This is because the deceleration parameter, $q_0$, is itself an order $1$ quantity. The back-reaction in this quantity is therefore still small compared to the corresponding FLRW value, even though its absolute magnitude has increased from the terms that enter into the Friedmann equations. At scales of about $100$ Mpc, we estimate that these corrections amount to changes at the level of about $1$ part in $10^{4}$ in the deceleration parameter. 

The value of $\mathcal{B}_{3}$ in the absence of radiation and a cosmological constant is negative, meaning that discretizing the matter in this way leads to a small increase in acceleration. This is no surprise, as back-reaction has already been shown to increase $\ddot{a}/a$ and decrease $\dot{a}^2/a^2$. As the value of $q_0$ is simply given by the ratio of these two quantities (with a minus sign), we have that both types of back-reaction contribute cumulatively to the acceleration measured by this dimensionless parameter.

It can be seen from Fig. \ref{fig8} that radiation increases the back-reaction that occurs in the deceleration parameter. Positive values of $\Lambda$ have a small effect on $\mathcal{B}_{3}$, even though it does not have a noticeable effect on $\mathcal{B}_{1}$ or $\mathcal{B}_{2}$. This is because, in Eq. (\ref{b3}), we find that $\Lambda$ enters into the background terms that multiply $\mathcal{B}_{1}$ and $\mathcal{B}_{2}$. Negative values of $\Lambda$ can make a more sizeable contribution to the back-reaction of $q_0$, and can even cause the back-reaction term to contribute to deceleration, if its magnitude is large enough. The effect of positive spatial curvature on $\mathcal{B}_{3}$ can also be large, but in this case causes extra acceleration. One should keep in mind, however, that for both of these last two cases the background value of the deceleration also diverges as  $\Omega_{\Lambda} \rightarrow - \Omega_{M}$ and $\Omega_{k} \rightarrow -\Omega_{M}$. Finally, and unlike in the acceleration and constraint equations, a negative value for the spatial curvature provides only a small correction to the value of $\mathcal{B}_{3}$.

The effects on $\mathcal{B}_{3}$ of simultaneously adding negative spatial curvature, positive cosmological constant, and non-zero radiation are displayed in Figs. \ref{fig10}-\ref{fig13}. It can be seen from Fig. \ref{fig10} that, in the presence of radiation, negative spatial curvature has only a small effect on the back-reaction. Similarly, in Fig. \ref{fig11}, it can be seen that positive values of $\Lambda$ have a small effect on the back-reaction term, when radiation is present. On the other hand, in Fig. \ref{fig13}, it can be seen that although positive $\Lambda$ and negative spatial curvature have only a small effect on the back-reaction in the absence of radiation, these effects are comparable to each other when radiation is absent. In this case, for small values of $\Lambda$, we have a small correction to the absolute value of $\mathcal{B}_{3}$, with a maximum at $\Omega_{\Lambda} = 0.5\Omega_{M}$. Negative spatial curvature does not affect $\mathcal{B}_{3}$ for small values of $\Lambda$, but does become increasingly significant as the value of $\Lambda$ increases.

\begin{figure}[b!]
\includegraphics[width=0.52 \textwidth]{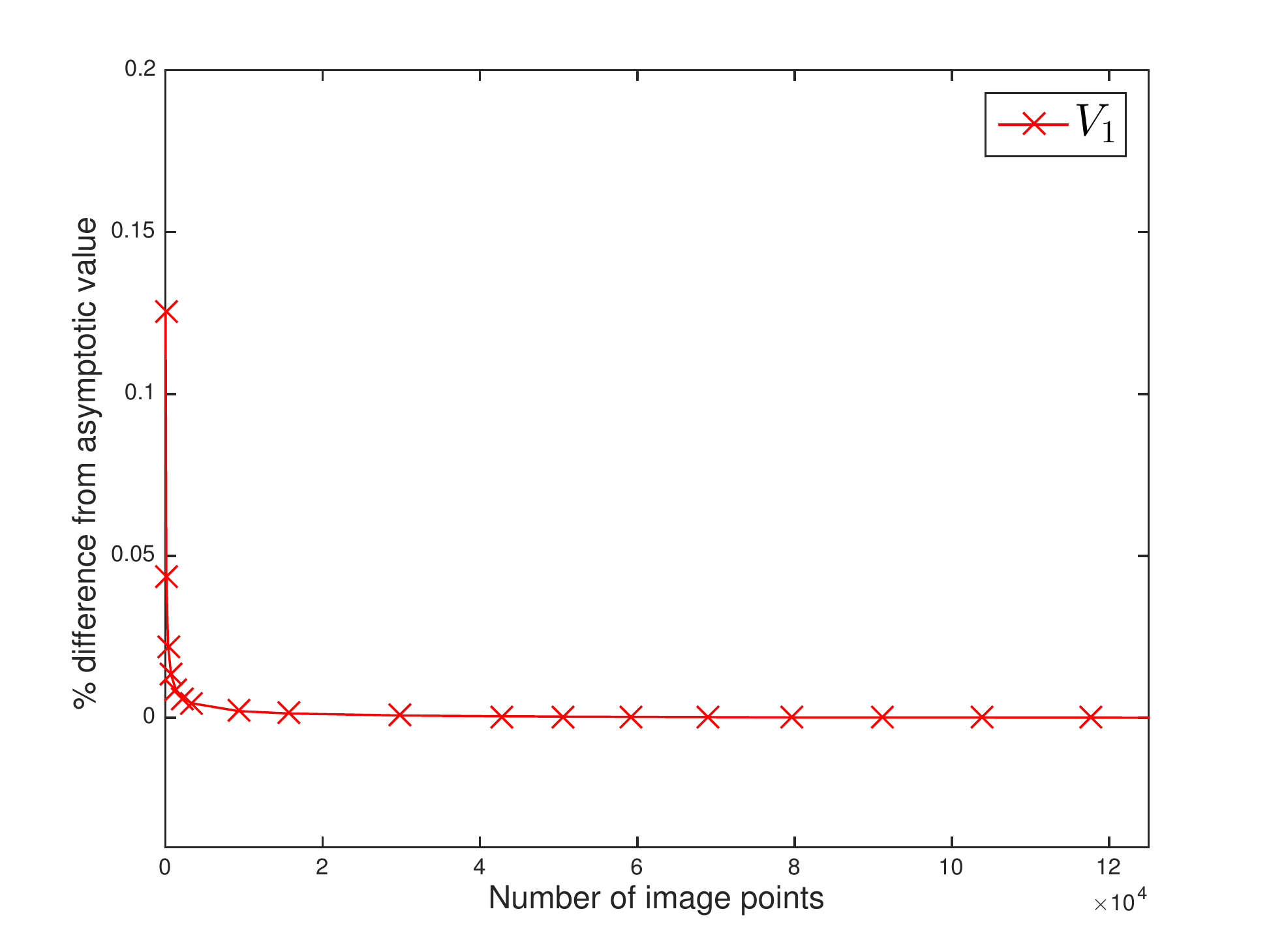}
\caption{\label{fig14} The percentage difference from the asymptotic value of $V_{1}$, for various different numbers of image points in the partial sum.}
\end{figure}

\section{CONCLUSIONS}

We have constructed an original framework that can be used to quantify the effects that radiation, spatial curvature and $\Lambda$ have on the cosmological back-reaction that results from the existence of non-linear inhomogeneities. Our approach is based on modelling the universe as a regular lattice, in which all structure is periodic. The geometry of space-time within each individual cell is then taken to be close to Minkowski space, and a post-Newtonian perturbative expansion is used to model all gravitational fields and matter content. By patching these cells together, using Israel's junction conditions at reflective symmetric boundaries, we finally construct a global and dynamical space-time. We derived an acceleration equation that describes the expansion of this emergent cosmology, and which is valid for any arbitrary distribution of matter within each cell (as long as it is distributed periodically). This equation is valid in the presence of both a barotropic fluid, with unspecified equation of state, and a cosmological constant.

Having derived the equations that govern the general case, we then simplified our equations by considering the specific example of a point-like mass at the centre of each lattice cell, in a sea of radiation and in the presence of a cosmological constant. The back-reaction terms generated by the matter fields alone behave like radiation in the Friedmann equation, as found in \cite{vaas}. The presence of actual radiation, however, reduces the magnitude of the back-reaction in both the acceleration and constraint equations. In contrast, we find that the cosmological constant has a negligible effect on back-reaction, and that spatial curvature can have a significant effect depending on whether the Universe is open or closed. These results explain why the leading-order effects of back-reaction occur at the level of linear-order perturbations in cosmological perturbation theory \cite{lam,radlam1,radlam2}, even though they require second-order gravity in order to be calculated.

In future work we will calculate observables in these models, by solving the equations that govern the expansion of a beam of light \cite{toapp2}. We also aim to further improve their realism by reducing the symmetries required at the junctions between cells.

\vspace{15pt}
\section*{ACKNOWLEDGEMENTS}
We are grateful to I. Brown, P. Carrilho, D. Dold, P. Fleury and S. Imrith for helpful discussions and comments. VAAS and TC both acknowledge support from the STFC.

\begin{table}[b!]
\begin{tabular}{ | c | l |  }
    \hline 
    \textbf{\, Constant \,} & \textbf{\, Asymptotic value \,} \\ \hline 
    $D$ & $\qquad \phantom{-}1.44 \ldots$   \\ \hline 
    $E$ & $\qquad \phantom{-}0.643 \dots$ \\  \hline
    $F$ & $\qquad -1.62 \dots$  \\ \hline
    $P$ & $\qquad \phantom{-}0.304 \dots$ \\ \hline
    $V_{1}$ & $\qquad \phantom{-}2.31 \dots$ \\ \hline 
\end{tabular}
\caption{\label{tab3} The numerical values of $D$, $E$, $F$, $P$, and $V_{1}$ that are approached as the number of reflections used in the method of images diverges to infinity.} 
\end{table}

\section*{Appendix: Numerical coefficients} \label{appA}

The numerical constants that appear in the acceleration equation \eqref{accmink} are given, in terms of the variables used in \cite{vaas}, by
\begin{align}
\mathcal{A}_{1} =& \frac{D}{3} - \frac{E}{2} + \frac{7\pi}{27} - \frac{F}{6} + \frac{P}{12} \, , \nonumber \\ \nonumber \\
\mathcal{A}_{2} =&  \frac{13\pi}{27} + \frac{16D}{3} - 4E - 8V_{1} - \frac{4F}{3}+ \frac{4P}{3} \, , \nonumber \\ \nonumber \\
\mathcal{A}_{3} =&\frac{5\pi}{216}  -\frac{2D}{3} - \frac{E}{2} + \frac{V_{1}}{3} - \frac{F}{6}-\frac{P}{6} \, , \nonumber \\ \nonumber \\
\mathcal{A}_{4} =& -\frac{5 \pi}{6} +\frac{F}{2} +\frac{3E}{2} \, .
\end{align}
The numerical values of $\mathcal{A}_{1}$, $\mathcal{A}_{2}$, $\mathcal{A}_{3}$ and $\mathcal{A}_{4}$ are given in Table \ref{tab1}, and the numerical values of $D$, $E$, $F$, $P$  and $V_{1}$ are given in Table \ref{tab3}. The quantity $V_1$, which is defined by
\begin{equation}
V_1 \equiv \frac{\int_{-X}^{X} \Phi_M dx dy dz}{4 G M X^2} \, ,
\end{equation}
converges to its limiting value quickly as the number of image masses is increased, as illustrated in Fig. \ref{fig14}. The convergence of $D$, $E$, $F$, $P$ and $V_{1}$ is given in \cite{vaas}.

\end{document}